\documentclass[showpacs,apl,twocolumn,floatfix]{revtex4}

\usepackage{graphicx,float,amsmath}
\usepackage{ mathrsfs }
\usepackage{natbib}

\begin{document}
\date{\today}

\title{Luminescence imaging of photoelectron spin precession during  drift in p-type GaAs}

\author{V.~Notot$^1$}
\author{D.~Paget$^1$}
\author{A. C. H. ~Rowe$^1$}
\author{L.~Martinelli$^1$}
\author{F.~Cadiz$^2$}
\author{S.~Arscott$^3$}

\affiliation{%
$^1$Physique de la mati\`ere condens\'ee, Ecole Polytechnique, CNRS, Universit\'e Paris Saclay, 91128 Palaiseau, France}

\affiliation{%
$^2$Universit\'e de Toulouse, INSA-CNRS-UPS, 31077 Toulouse Cedex, France}

\affiliation{%
$^3$Institut d'Electronique, de Micro\'electronique et de Nanotechnologie (IEMN), University of Lille, CNRS, Avenue Poincar\'e, Cit\'e Scientifique, 59652 Villeneuve d'Ascq, France}

\begin{abstract}
Using a microfabricated, p-type GaAs Hall bar, is it shown that the combined application of co-planar electric and magnetic fields enables the observation at 50 K of spatial oscillations of the photoluminescence circular polarization due to the precession of drifting spin-polarized photoelectrons. Observation of these oscillations as a function of electric field $E$ gives a direct measurement of the minority carrier drift mobility and reveals that, for $E=800$ V/cm, spin coherence is preserved over a length as large as  $25 \mu$m.  
 
\end{abstract}
\maketitle

In the context of future active spintronic devices, the diffusion \cite{volkl2011, cadiz2014} and drift \cite{kato2003, Huang2008, kikkawa1999, hernandez2016, yu2002, kotissek2007} of spins in semiconductors have been investigated by numerous authors.  N-type GaAs seems particularly promising because of the weak spin relaxation \cite{dzhioev2002}, but p-type material cannot be overlooked since in proposed bi-polar spintronic devices \cite{zutic2007}, it is the minority carrier (electron) spin that determines the common base current gain. While in p-type material charge transport has been investigated \cite{luber2006, ito1989, cadiz2015}, very few studies have considered spin transport \cite{cadiz2015d, cadiz2015b}. Time-resolved investigations have been used \cite{henn2013} but these investigations do not provide a direct spatial imaging. On the other hand, continuous wave (CW) imaging of spin transport, using luminescence  \cite{favorskiy2010}, or Kerr microscopy \cite{volkl2011}  gives diffusion and drift lengths from which it is possible to obtain $\mu \tau$ for the charge transport where $\mu$ is the mobility and $\tau$ is the lifetime, or $\mu \tau_s$ for spin transport where $\tau_s$ is the spin lifetime. Consequently, CW determination of mobilities or of diffusion constants requires separate determinations of lifetimes \cite{cadiz2014}.\ 

\begin{figure}[tbp]
\includegraphics[clip,width=7 cm] {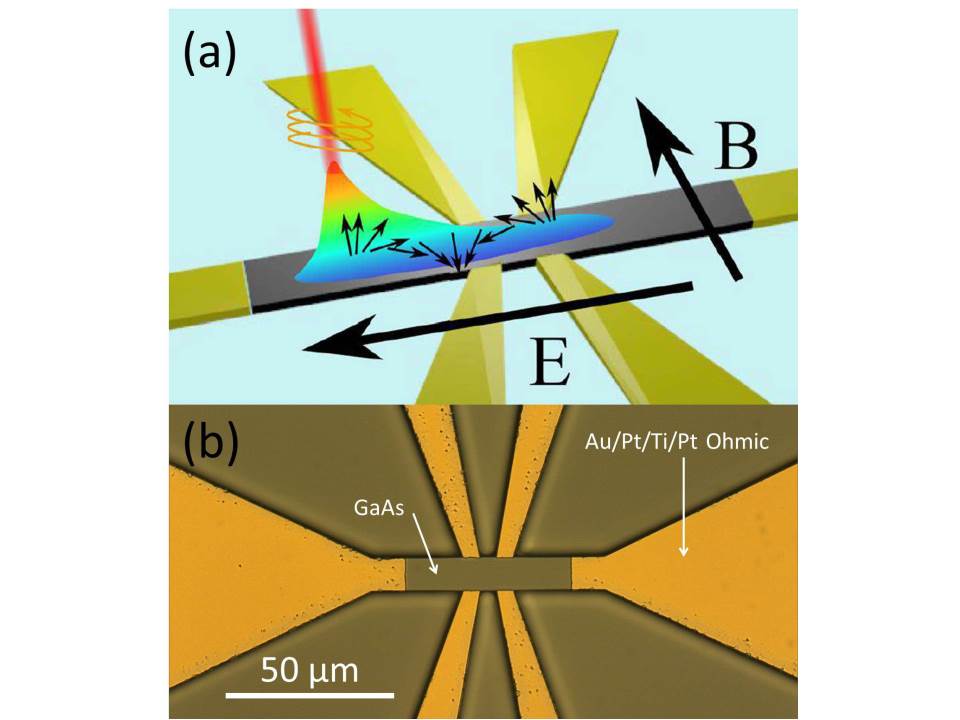}
\caption{Panel a shows the principle of the experiment. Photoelectrons are created by a circularly-polarized tightly-focussed light excitation in a Hall bar where an electric field is applied. The luminescence degree of circular polarization of photoelectrons drifting in this electric field is imaged. This imaging reveals the precession of the photoelectronic spins in a magnetic field applied in an arbitrary direction parallel to the sample plane. Panel b shows an optical microscope image of the microfabricated Hall bar. }
\label{Fig01}
\end{figure}

In the present work, we investigate spin transport in p-type GaAs using a CW technique in which precession in a magnetic field tranverse to the light excitation acts to effectively time-resolve the experiment so that the relevant parameters for spin drift transport can be determined.  As shown in  Fig. \ref{Fig01}, in the same way as performed elsewhere using Kerr microscopy \cite{crooker2005, furis2007}, we use a microfabricated Hall bar, fabricated from a 3 $\mu$m -thick p-type GaAs sample (acceptor doping range  $10^{18}$ cm$^{-3}$), with the long axis along which the electric field is applied parallel to  the $[110]$ crystallographic direction \cite{cadiz2015c}.  The maximum value of of the electric field used here (800 V/cm) is well below that required to saturate the drift velocity at low temperatures \cite{elela2011}. The photoelectrons are generated by a tightly-focussed 12 $\mu$W  CW excitation by a laser at 1.59 eV, so that transport is not affected by ambipolar effects \cite{cadiz2015c, cadiz2015d}, or by Pauli blockade \cite{cadiz_prl2013, cadiz2015b}. During electron drift in the  electric field $E$, their spin precesses in a magnetic field $B$ (0.23 T) applied in the sample plane. Spin precession results in coherent spatial oscillations of the spin density, which are observed by monitoring the degree of circular polarization of the luminesence in the direction parallel to the photo-excitation wavevector. Qualitatively, the spatial period $l $ of the oscillations is related to the photoelectron drift speed  $\vec v=\mu_e \vec E$ , where $\mu_e$ is the mobility, and to the precession frequency in the magnetic field $\omega = \hbar ^{-1} g^{\ast}\mu_{B}B$, where $g^{\ast}=-0.4 $ is the effective Land\'e factor, $2 \pi \hbar$  is Planck's constant and $ \mu_{B}$ is the Bohr magneton. One has
 \begin{align}
	l \approx \mu_{e}\frac{E}{B} \frac{h}{g^{\ast}\mu_{B}} 
	\label{eq_mu_elec}
\end{align}

\noindent
Thus, the measurement of $l$ directly gives the photoelectron drift mobility, while analysis of the damping gives information on the mechanisms for loss of spatial spin coherence, which are mostly due to spin relaxation as characterized by the time $T_1$. As will be discussed here, this is an advantage of the polarized luminescence technique when compared to Kerr imaging where the damping is determined by the spin lifetime, $1/\tau_s = 1/\tau + 1/T_1$\cite{crooker2005, furis2007}. When spin relaxation times are equal to or longer than minority carrier lifetimes, the damping of spin precession oscillations is therefore stronger in Kerr microscopy. Precession can thus be observed over longer distances using polarized microluminescence.\ 

\begin{figure}[tbp]
\includegraphics[clip,width=9 cm] {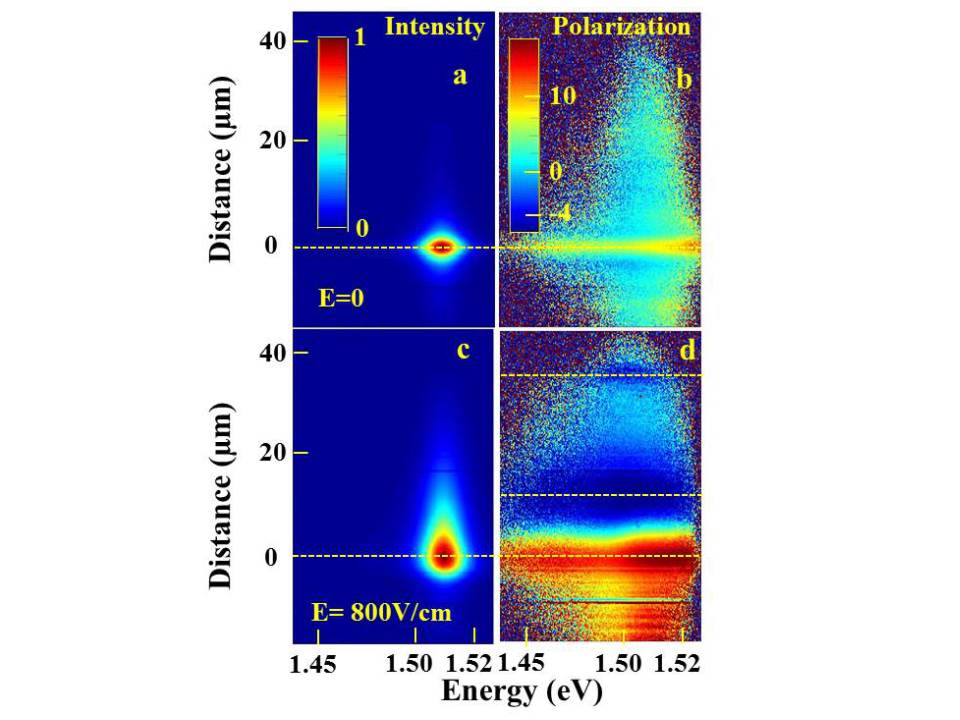}
\caption{Combined effects of electric and magnetic fields on the images detected by the CCD Camera at the exit of the spectrometer. The temperature is 50K and a magnetic field of 0.23T is applied perpendicular to the excitation light direction. The vertical direction represents the distance on the sample to the excitation spot, while section of the images along the horizontal direction shows an intensity (panels a and c) and a circular polarization spectrum (Panels b and d) at the corresponding position at the sample surface. Panels a and c represent the intensity images for $E=0$ and $E=800$ V/cm and their comparison reveals the drift of the electrons in the electric field. Panels b and d represent the corresponding images of the luminescence degree of circular polarization and show the oscillations induced by precession during drift.}
\label{Fig02}
\end{figure} 

The luminescence is focused on the entrance slit of the spectrometer and one monitors the image provided by a CCD camera placed in the output plane. For $\vec{ E} = \vec{0}$, this image is shown in Panel a of  Fig. \ref{Fig02}. A cut of this image along its horizontal axis, perpendicular to the entrance slit, gives the local luminescence spectrum at a given distance from the excitation spot and reveals the usual emission of $p^+$ GaAs \cite{feng1995}. Conversely, a cut of the image along the vertical axis (parallel to the entrance slit) gives the spatial intensity profile at a given energy, along a line on the sample parallel to the spectrometer entrance slit and to the electric field (the origin of ordinates denoting the distance to the excitation spot). Panel c of Fig. \ref{Fig02} shows the corresponding image at $E=800$ V/cm. This image reveals a tail of drifting electrons extending over several tens of microns that is essentially independent of energy. 

\begin{figure}[tbp]
\includegraphics[clip,width=7 cm] {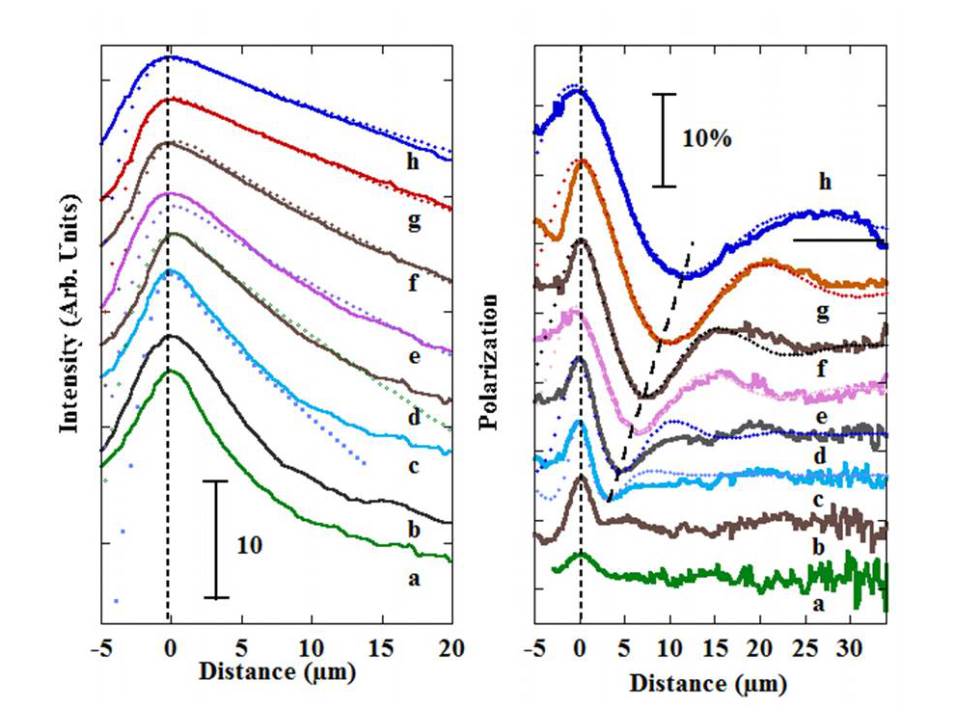}
\caption{The left panel shows the spatial profiles, shifted vertically for clarity, of the luminescence intensity (left panel) and polarization (right panel), for a zero electric field (a) and for $E$ equal to 50V/cm (b), 100 V/cm (c), 200 V/cm (d), 300V/cm (e), 400 V/cm (f), 600V/cm (g) and 800V/cm (h). The magnetic field value is 0.23 T. Spatial oscillations of the polarization due to electronic spin precession during drift appear for a field larger than 100V/cm and, as expected from Eq. (\ref{eq_mu_elec}), their period increases with electric field. Also shown by dotted lines in the two panels are calculations in the drift regime using parameter values given by Table \ref{table1}.}
\label{Fig03}
\end{figure}

Liquid crystal modulators are also used to control the helicity of the excitation and to select the $\sigma ^{\pm}$-polarized components of the luminescence, of intensity $I(\sigma ^{\pm})$. It is then possible to monitor the difference signal $ I_{D}= I(\sigma ^+) - I(\sigma ^-)$, in order to obtain the polarization image $\mathscr{P} = [I(\sigma ^+) - I(\sigma ^-)]/ I $. The image of this ratio taken at $\vec{ E} = \vec{0}$ is shown in Panel b of  Fig. \ref{Fig02}. Note that the polarization is higher at 1.52 eV than at 1.51 eV corresponding to the maximum of the luminescence emission. This reveals that hot electrons are more polarized than electrons at the bottom of the conduction band.

Panel d of Fig. \ref{Fig02} shows the polarization image for $E=800$ V/cm. The polarization peak, in the same way as Panel b, corresponds to hot electrons, and its amplitude is increased to about 15 $\%$ because of the reduced electronic lifetime at the  excitation spot. The key result is the appearance of spatial oscillations with regions of negative  polarization, shown by dashed lines at distances of 10 $\mu$m and 27 $\mu$m, separated by positive regions. Fig. \ref{Fig03} shows the spatial profiles of the intensity (left panel) and polarization (right panel) at the energy of the maximum luminescence signal, for the various electric field values.  The polarization oscillations are observed for an electric field larger than $50$ V/cm and, as expected, their spatial period increases with $E$. For lower electric fields, the dominant transport process is spatially incoherent diffusion so that no oscillations are visible. 

We now calculate the photoelectron concentration in steady-state. This concentration is given by the drift-diffusion equation
\begin{align}
0= g(r,z) -\frac{n}{\tau}   + \vec \nabla  (\mu_e n \vec E+ D \vec \nabla n), 
\label{eqcharge}
\end{align}
where  $D$ the diffusion constant of photoelectrons. The generation rate $g(r,z)$ depends on depth $z$ and distance to the center of the excitation spot $r$ according to  $g(r,z)= \alpha g_0 \exp[-\alpha z - (r/\sigma)^2]$, where $\alpha$ is the light absorption coefficient, and $\sigma$ is the gaussian half width of the excitation spot.  To calculate the spatial distribution of the spin density, it is convenient to define the complex spin density $s_{\pm}= s_z \pm i s_Y$ where $Y$ is the direection of the sample plane perpendicular to the orientation of the magnetic field, which can be arbitrary in the sample plane. The drift-diffusion-precession equation for $s_{\pm}$ is 

\begin{align}
 	g(r, z) \eta \mathscr {P}_i - \frac{s_{\pm}}{\tau_{s}}(1 \mp i \frac{B}{\Delta B}) + \overrightarrow{\nabla} \cdot [\mu_e \overrightarrow{E}  s_{\pm} + D\overrightarrow{\nabla}s_{\pm}] = 0
 \label{eqIspin}
 \end{align}
     This equation is obtained from the charge equation by replacing $g(r,z)$ by $g(r, z) \eta \mathscr {P}_i$, where $\mathscr {P}_i$ is equal to $\pm 0.5$ for $\sigma^{\mp}$-polarized light excitation, and $\eta$ accounts for polarization losses during thermalization.  Here  $\tau$ has been replaced by the complex spin lifetime $\tau_s/(1 \mp i  B/\Delta B)$, where $\Delta B= \hbar/(g^{\ast}\mu_{B} \tau_s)$ is the Hanle linewidth. \

A solution of these equations for a film of thickness $d$ is given in the Supplementary information. We consider here the limit where $SD/d<<1$ and $S'D/d<<1$, where $S$ and $S'$ are  the recombination velocities of the front and back surfaces. The spatial dependence of the degree of circular polarization $\mathscr{P}_i s/n$ of the luminescence, after averaging over the sample thickness, is dominated by the spatial mode of lowest order and given by 
\begin{align}
I (r)= A \frac{\alpha g_0 \tau e^{\beta x}}{2 \pi}  (\frac{\mathscr{L}_1 }{L_{d}^{*}})^{2}   \Big( K_0 (\frac{\lvert r \rvert}{\mathscr{L}_1}) * e^{[-(\frac{r}{\sigma})^2 +\beta x]}  \Big) , 
\label{eqIsum1}
\end{align}
where A is a constant and $K_0$ is the  modified Bessel function of the second kind, the symbol $*$ stands for two-dimensional convolution and the effective length  $\mathscr{L}_1$ is given by 

\begin{align}
\frac{1}{\mathscr{L}_1}=  \sqrt{ \frac{1}{L_{d}^{*2}}  +\beta ^2}   , 
\label{eqL1}
\end{align}

\noindent
where $L_{d}^{*}= \sqrt{D\tau^{*}}$,  and $1/\tau^{*}= 1/\tau +(S+S' )/(Dd)$. The inverse length $\beta$ is given by  
\begin{align}
	\beta= - \frac{\mu E}{2D}=-\frac{q E}{2k_BT_{e}},
	\label{eq_xi}
\end{align}
where $q$ is the absolute value of the electronic charge and $k_B$ is Boltzmann's constant. As shown by the right hand side of Eq. (\ref{eq_xi}) where Einstein's relation has been applied, $\beta$ only depends on $E$ and on the  temperature $T_e$ of the photoelectron gas. At large charge drift distances $x$  parallel to the electric field  and  in the charge drift regime defined by $\beta L_d^{*}>>1$, the profile described by Eq. (\ref{eqIsum1}) decays like $\exp[-x/(\mu E \tau^{*})]$. The spatial profile of the polarization is given by 
\begin{align}
	   \mathscr{P}(r) = & \eta \mathscr{P}_i^2 \frac{\tau_s (1+\beta^2 L_{d}^{*2}) }{\tau (1+\beta^2 L_{s}^{*2})} \nonumber \\ & 
  Re \Bigg\{  \frac{  K_0 (\frac{\lvert r \rvert}{\mathscr{L}_1^s}) * e^{[-(\frac{r}{\sigma})^2 +\beta x]} }
	 { (1 \mp iB /\Delta B^*)  \Big( K_0 (\frac{\lvert r \rvert}{\mathscr{L}_1}) * e^{[-(\frac{r}{\sigma})^2 +\beta x]}  \Big)}
  \Bigg\}	 	
	\label{eqpol}
\end{align}
\noindent
where  $\Delta B^* =\Delta B (1+\beta^2 L_{s}^{2})$, $L_{s}^{*}= \sqrt{D\tau_s^{*}}$ and $1/\tau_s^{*}= 1/\tau_s +(S+S' )/(Dd)$, and $\mathscr{L}_{1s}$ is given by
\begin{align}
\frac{1}{\mathscr{L}_{1s}}=  \sqrt{ \frac{1 \mp iB/\Delta B}{L_{s}^{*2} }  +\beta^2 }, 
\label{eqL1s}
\end{align}. 

\noindent
Because $\mathscr{L}_1^s $ is complex,  the function $K_0$ in the numerator of Eq. (\ref{eqpol}) has an oscillating spatial behavior. These oscillations are superimposed on a spatial polarization decay, which is described by $\exp[-x/(\mu E T_{1} )]$ at large distance and in the spin drift regime defined by $\beta L_d^{*}>>1$ and $\beta L_s^{*}>>1$. In contrast, the decay of the spin density $s$ in the same conditions, probed by techniques such as Kerr imaging, is described by $exp[-x/(\mu E \tau_s^{*})]$ and can be significantly faster.  

\begin{figure}[tbp]
\includegraphics[clip,width=6 cm] {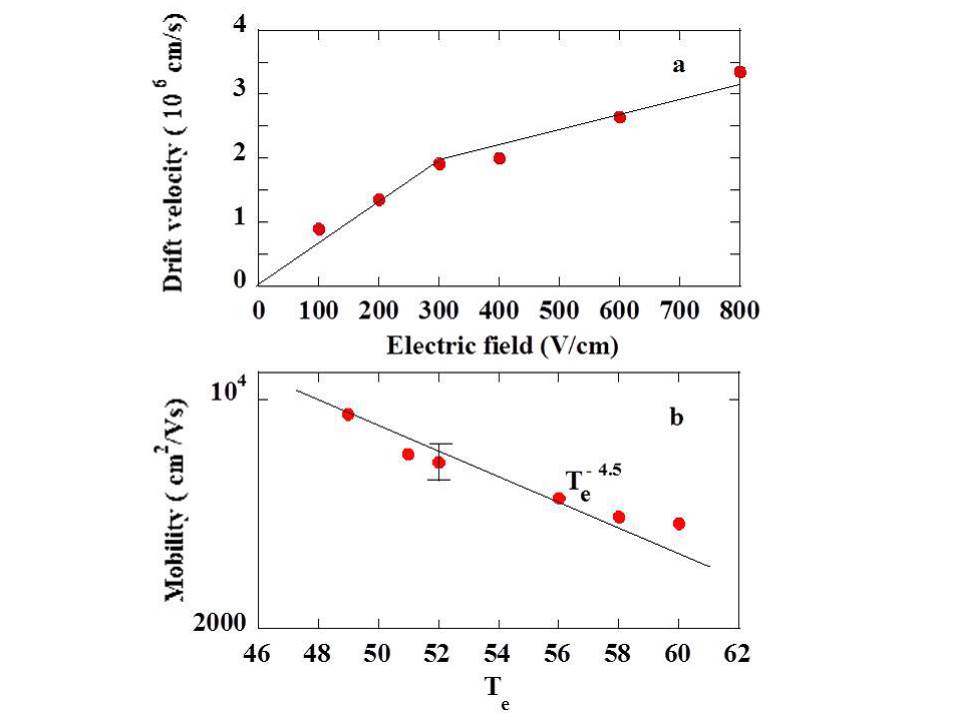}
\caption{Panel a shows the electric field dependence of the measured photoelectron drift speed $v=\mu E$. Panel b shows the dependence of the mobility as a function of electronic temperature. One finds a power law of exponent -4.5, similar to the result found elsewhere \cite{cadiz2015}.}
\label{Fig04}
\end{figure}

The use of Eq. (\ref{eqpol}) for analyzing the spatial oscillations directly gives the photoelectron mobilities, since the  positions  of the extrema do not depend on other parameters such as the charge and spin lifetimes. The  mobility values, given in Table \ref{table1}, strongly decrease with increasing electric field. As shown elsewhere \cite{cadiz2015}, this effect reveals a dependence of the mobility on electronic temperature. Shown in  panel b of Fig. \ref{Fig04} is the dependence of $\mu_e$ as a function of $T_e$, which is obtained from the slope of the high energy tail of the luminescence spectrum. One sees that the mobility decreases with $T_e$ as a power law of exponent -4.5, that is, very close to that found in Ref. \cite{cadiz2015}. Shown in the panel a of Fig. \ref{Fig04} is the dependence of the drift velocity $\mu_e E$ as a function of electric field. For electric fields lower than about $300$ V/cm, the velocity is approximately proportional to the electric  field, suggesting quasi ohmic transport. The velocity slightly saturates for higher fields, with a maximum value, of several $10^6$ cm/s at $800$V/cm, which is smaller than the saturation velocity by at least one order of magnitude \cite{elela2011}. 

Note also that, as shown in Table \ref{table1}, the polarization at $r=0$ strongly increases with the electric field and reaches  its maximum value $\eta \mathscr{P}_i^2$ for $E=800$ V/cm. This is because in the high electric field regime, such that $\beta  L_{d}>>1$ and $\beta L_{s}>>1$, the effective time of residence at the excitation spot (half-width $\sigma$) is determined by drift [$\tau _{drift} = \sigma /(\mu E) \approx 10$ ps], which is much smaller than the times for spin relaxation, precession, and recombination. Conversely, at E=0, the effective spin lifetime  at the excitation spot is determined by diffusion ($\tau _{dif} \approx \sigma ^2/D \approx 60$ ps) and is small with respect to  the inverse precession frequency in the magnetic field ($\omega^{-1} \approx 10^{-10}$ s/rd)  so that the polarization value is strongly affected by the Hanle effect. 

\begin{table}[tp]\footnotesize
\caption{Analysis of intensity and polarization profiles in the drift regime ($E>50 $ V/cm). Measured values of electronic temperature and polarization at the excitation spot, and values of electron mobility,  $\eta \mathscr{P}_i^2$, and spin and charge lifetimes used in the adjustments of the curves in Fig. \ref{Fig03}. Numbers in parentheses give the uncertainties of the determinations}
\label{table1}\centering 
\begin{tabular}{cccccccc}

E & $T_e$ (K)&$\mathscr{P}(0)$  & $\mu _e$    & $\eta \mathscr{P}_i^2$ & $T_1$      & $\tau ^*$  & $\mu E T_1 $    \\ \hline
(V/cm) & (K) &($\%$)            & (cm$^2$/V.s)&  ($\%$)                & (ns)       & (ns)       & ($\mu$m)   \\ \hline
100    & 49  & 6.4              & 9000        & 19                     & 0.32(10)   & 0.33(5)    & 3.0\\ \hline
200    & 51  & 8.8              & 6800        & 13.7                   & 0.50(10)   & 0.33(5)    & 7.0 \\ \hline
300    & 52  & 9                & 6400        & 11.5                   & 0.51(10)   & 0.33(5)    & 9.8 \\ \hline
400    & 56  & 12.6             & 5000        & 14                     & 0.47(10)   & 0.33(5)    & 9.4  \\ \hline
600    & 58  & 14.8             & 4400        & 15                     & 0.47(10)   & 0.33(5)    & 12.4  \\ \hline
800    & 60  & 16.4             & 4200        & 17                     & 0.33(10)   & 0.33(5)    & 11  \\ \hline
\end{tabular}
\end{table}

	For $E>50$ V/cm, once the mobility is determined, it is possible to interpret the spatial luminescence and polarization profiles using parameter values shown in  Table \ref{table1}. As seen in the left panel of Fig. \ref{Fig03}, the intensity profiles are correctly interpreted by using a value  of $\tau^*=$0.33 $\pm$ 0.05 independent on $E$. While the uncertainty of this determination masks the possible increase of $\tau^*$ with $T_e$ \cite{note4}, this value lies in the $0.32-0.40$ ns range found for $E=0$ in the same temperature range using time-resolved luminescence \cite{cadiz2014}. The value of $T_1$, determined from the damping of the polarization decay is approximately independent of electric field and is of the order of $0.45 \pm 0.1$ ns. Again, this value is comparable with the zero-field value of $0.5$ ns found elsewhere in the same temperature range \cite{cadiz2014}. Finally, $\eta \mathscr{P}_i^2$ is estimated from $\mathscr{P}(0)$. Its value is consistently smaller than $\mathscr{P}_i^2 = 25\%$, because of polarization losses during thermalization. Note that the value of the spin coherence length $\mu E T_1$, defined as the characteristic exponential decay length of the polarization at large distance, is also given in Table \ref{table1}. As expected, this value increases with electric field and reaches a maximum value larger than $10$ $\mu$m at high electric field. This result is in agreement with the observation of a maximum in the polarization at a distance of $25 \mu$m, in Curve h of  the right panel of Fig. \ref{Fig03}, and shows that spin coherence is preserved after this relatively large distance and after a spin precession angle of $2\pi$.\
		
	In conclusion, we have investigated spin transport in p-type GaAs and have shown that application of an electric field is able to preserve spin coherence lengths up to about $25 \mu$m. This has been performed by imaging the spin precession in a magnetic field applied in the sample plane.  It is pointed out that the method described by Fig. \ref{Fig01} has further potentialities. Firstly, it is straightforward to analyze the spatial oscillations as a function of energy in the images of  Fig. \ref{Fig02} and therefore to investigate the value of the exponent $p$ of the dependence of  mobility on kinetic energy $\epsilon$ in the conduction band, defined by $\mu \approx \epsilon ^p$. As shown in the supplementary information, it is found that the oscillations do not depend on kinetic energy so that $p \approx 0$, in agreement with the screening by holes of photoelectron scattering mechanisms at this large doping, and with independent Hall effect measurements on this sample \cite{cadiz2015}.  Secondly, in the absence of an applied magnetic field, and for an appropriate drift crystallographic orientation, this same method may also be used  to probe effects of the spin-orbit interaction \cite{note3}.\\	 
\acknowledgements{The help of A. Wack, D. Lenoir and P. Njock for mounting  the $\mu$PL experimental setup is acknowledged. This work was partly supported by the French RENATECH network.}
\appendix

\bibliographystyle{apsrev}


\begin{thebibliography}{25}
\bibitem[{not({\natexlab{a}})}]{note4}
\bibinfo{note}{Since the drift velocity shown in Panel a of Fig. \ref{Fig04} is
  smaller than the thermal velocity of the photoelectron gas at $T_e =60$K by
  about one order of magnitude, it is considered that the photoelectron
  statistics are essentially described by the temperature $T_e$ of the
  photoelectron gas.}
\bibitem[{not({\natexlab{b}})}]{note3}
\bibinfo{note}{Here, we calculate that the Dresselhaus spin-orbit interaction,
  equivalent to a magnetic field of 0.05 T for the maximum electric field
  value, is too weak to induce observable spatial oscillations of the
  polarization.}
\end{thebibliography}

\end{document}